\documentclass[a4paper]{article}
\usepackage{amsmath,amsfonts,epsfig,psfrag}

\newcommand{\nn}{\nonumber}

\newcommand{\Lam}{\Lambda}
\newcommand{\lam}{\lambda}
\newcommand{\del}{\delta}
\newcommand{\Om}{\Omega}

\newcommand{\phbar}{{\phi}}

\newcommand{\Hbar}{{H}}
\newcommand{\Lbar}{{\Lambda}}
\newcommand{\diff}[2]{\frac{d #1}{d #2}}
\newcommand{\cdt}{\mathsf{t}}
\newcommand{\cdr}{\mathsf{r}}
\newcommand{\cdx}{\mathsf{x}}
\newcommand{\cdy}{\mathsf{y}}
\newcommand{\cdz}{\mathsf{z}}

\newcommand{\clp}{{\mathcal{P}}}
\newcommand{\clr}{{\mathcal{R}}}

\newcommand{\Rl}{\mathbb{R}}

\begin{document}

\noindent
{\bf\Large \textsf{Conformal models of de Sitter space, initial \\
conditions for inflation and the CMB}}

\vspace{0.4cm}

\noindent
{\large Anthony Lasenby\footnote{e-mail: \texttt{a.n.lasenby@mrao.cam.ac.uk}}
and 
Chris Doran\footnote{e-mail: \texttt{c.Doran@mrao.cam.ac.uk}}}

\vspace{0.4cm}

\noindent
Astrophysics Group, Cavendish Laboratory, Madingley Road, \\
Cambridge CB3 0HE, UK.

\vspace{0.4cm}

\begin{center}
\begin{abstract}
Conformal embedding of closed-universe models in a de Sitter
background suggests a quantisation condition on the available
conformal time.  This condition implies that the universe is
closed at no greater than the 10\% level.  When a massive scalar
field is introduced to drive an inflationary phase this figure is
reduced to closure at nearer the 1\% level.  In order to enforce
the constraint on the available conformal time we need to
consider conditions in the universe before the onset of
inflation.  A formal series around the initial singularity is
constructed, which rests on a pair of dimensionless,
scale-invariant parameters.  For physically-acceptable models we
find that both parameters are of order unity, so no fine tuning
is required, except in the mass of the scalar field.  For typical
values of the input parameters we predict the observed values of
the cosmological parameters, including the magnitude of the
cosmological constant.  The model produces a very good fit to the
most recent CMBR data, predicting a low-$\ell$ fall-off in the
CMB power spectrum consistent with that observed by WMAP.
\end{abstract}
\end{center}

\section{Introduction}

Experimental evidence is now strongly in favour of the idea that a
non-zero cosmological constant, or some form of `dark energy', is
currently responsible for around 70\% of the total energy density of
the universe~\cite{WMAP}.  There have been many theoretical attempts
to justify the reintroduction of a $\Lam$-term, something that was
once viewed as a rather ugly and unnecessary extension to classical
general relativity.  Many of these are based on ideas from particle
physics, including concepts such as spontaneous symmetry breaking and
false vacua.  These ideas can successfully explain the size of the
cosmological `constant' required during an inflationary phase, but all
such models have great difficulty in explaining the current scale of the
cosmological constant without implausible levels of fine tuning.

It could be argued that the current scale of the cosmological constant
suggests that its role may be more geometric, rather than
field-theoretic.  This idea fits in well with the gauge-theoretic
viewpoint of gravity that we have developed
elsewhere~\cite{DGL98-grav,gap}.  In the absence of any matter, a
non-zero cosmological constant implies that the universe should be
described by a de Sitter space.  This space should then form the
background for the gauge theory treatment of gravity, though the
details of such a theory remain to be fully worked out.

Here we explore some simpler consequences of the view that $\Lam$
should be viewed as a genuine \textit{constant} with a geometric
role to play.  The ideas outlined here are discussed in more
detail in~\cite{DL03-deS}.  We first introduce a new picture of
de Sitter geometry, which generalises the Poincar\'{e} disk
picture of hyperbolic geometry~\cite{gap,brannan-cup}.  Since all
cosmological models are conformally equivalent (they all have
vanishing Weyl curvature), all cosmologies with a de Sitter end
state will have a standard embedding in the conformal picture.
These embedding-s are revealing, and suggest a `preferred'
cosmology --- one with closed spatial sections, in which the
total available conformal time is $\pi/2$.  Such a model places
the initial singularity symmetrically in the centre of the de
Sitter picture. The requirement on the total conformal time
suggests the operation of a quantisation condition, and we
speculate as to how such a condition may arise. Of course, there
are many theoretical reasons for preferring spatial flatness
which is, after all, supposed to be one of the main predictions
of inflation. For example, there are difficulties in constructing
a homogeneous stress-energy tensor for Dirac fermions in anything
other than a spatially-flat universe~\cite{ish74,DGL98-grav}. But
closed universe models have their own attractive features,
particularly in their inherent finiteness.  Furthermore, in
closed (or open) cosmologies the scale factor can be determined
absolutely from the equation
\begin{equation}
\frac{1}{R^2} = \frac{8 \pi G}{3} \rho -H^2 + \frac{\Lam}{3}.
\label{curv}
\end{equation}
This result simplifies calculations for perturbations, as we do not
have to track the scale factor through reheating in order to compare
physical scales today with those during inflation.

Several other authors have of course been interested recently in
closed universe models, motivated partly by the apparent
low-$\ell$ turn down in the WMAP results~\cite{WMAP}. For
example, Efstathiou~\cite{efs03} proposed a phenomological model
in which the low-$k$ primordial spectrum in a closed model has an
exponential cutoff on a scale comparable to the curvature scale.
On the other hand, Starobinsky~\cite{sta96}, predicted that the
effect of closure for adiabatic scalar perturbations was in
general to increase low-$\ell$ values.  This is clearly
an area where more detailed calculations are necessary.

On the observational side, Efstathiou (e.g. \cite{efs04}) and
others have queried whether foreground effects on power spectrum
estimation could be affecting the low-$\ell$ modes. However, it
is likely that the low value seen for the quadrupole mode is
still statistically significant, and thus represents something
which needs explaining in the standard model. Other explanations
which have been given include \cite{cont03} which has a period of
fast-roll in a flat model, and also topological attempts at
explanation, via effects in a compact closed
universe~\cite{uzan04}.

What we will look at here is simple closed universe, with no
topological effects.  But, by pursuing our boundary condition, we will
find strong constraints among cosmological parameters. In particular,
for a cosmological model with a given equation of state, the condition
that the total elapsed conformal time is equal to $\pi/2$ singles out
a unique trajectory in the $(\Om_M, \Om_\Lam)$ plane. For example, a
simple dust model predicts a universe that is closed at around the
10\% level. This represents an upper limit, and both inflationary and
radiation-dominated epochs drive this figure down.  A prediction of
the cosmological model can equally be interpreted as a prediction of
$\Lambda$ and, for typical input parameters, we find that
\begin{equation}
\Lam \approx \exp(-6N) l_{p}^{-2}. \label{eqn:for-Lambda}
\end{equation}
Here $N$ is the number of e-foldings in the inflationary region, and
turns out to be roughly 46, putting $\Lam$ in precisely the observed
range.

The model of inflation we consider is the simplest available ---
that of a massive scalar field.  This produces an extremely tight
model that agrees well with all current experimental data.  In
order to apply our boundary condition we need to study the
evolution of the scalar field from the initial singularity,
through the inflationary region, before matching onto a standard
cosmological model.  It has become quite common in the
inflationary community to ignore the initial singularity and
concentrate instead on suitable initial conditions for inflation.
It might have been hoped that the peculiar nature of the
inflationary stress-energy tensor, which violates the weak energy
condition, might circumvent some of the standard singularity
theorems.  But this is not the case.  Inflationary models are
geodesically incomplete in the past, and the cosmological
equations can be easily run backwards in time to reveal the
initial singularity~\cite{brand01,borde03}.  In this respect,
inflation has little to say about the `specialness' of the
initial singularity (zero Weyl curvature) which can be regarded
as one of the outstanding problems in physics~\cite{pen-road}.

Expanding the field equations around the initial singularity can
be performed using an iterative scheme in powers of $t$ and $\ln
t$.  The resulting series are governed by two parameters, which
effectively control the degree of inflation and the curvature.
Fixing both of these parameters to be of order unity produces
inflationary models in a closed universe which are consistent
with observation.  This appears to provide a counterexample to
statements that it is difficult to obtain closed universe
inflation without excessive fine
tuning~\cite{lin03,uza03,ellis02a,ellis02}.

As the universe exits the inflationary region it evolves as if it
had started from an effective big-bang, with a displaced time
coordinate. Photons have travelled an appreciable distance by the
end of inflation, which alters how we apply the $\pi/2$ boundary
condition. The result of these effects is the imposition of a
see-saw mechanism linking the current state of the universe and
the initial conditions. The more we increase the number of
e-folds during inflation, the smaller the value of the
cosmological constant, and vice-versa.  With initial conditions
chosen to give the required number of e-foldings to generate the
observed perturbation spectrum, we find that the predicted
universe is closed at the level of a few percent, in excellent
agreement with observation.  More detailed calculation also
reveals a dip in the low-$\ell$ part of the power spectrum.
However, these calculations are difficult and we comment on the
problems that must be overcome in order to make this prediction
robust.

Unless stated otherwise we work in units where $G=c=\hbar=1$.  Where it
adds clarity, factors of $G$ are included, so that $G$ has
dimensions of $(\mbox{distance})^2$.

\section{Conformal pictures of de Sitter space}

\begin{figure}\begin{center}
\includegraphics[height=6cm]{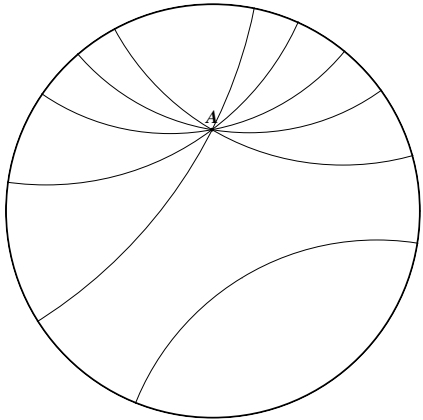}
\caption{\textit{The Poincar\'{e} disc.}
Points inside the disc represent points in a 2-dimensional
non-Euclidean (hyperbolic) space.  A set
of $d$-lines are also shown.  These are (Euclidean) circles that
intersect the unit circle at right angles.  Given a $d$-line and point
$A$ not on the line, one can find an infinite number of lines through
$A$ that do not intersect the line.}
\label{fig2}
\end{center}\end{figure}

\begin{figure}\begin{center}
\includegraphics[height=5cm,angle=-90]{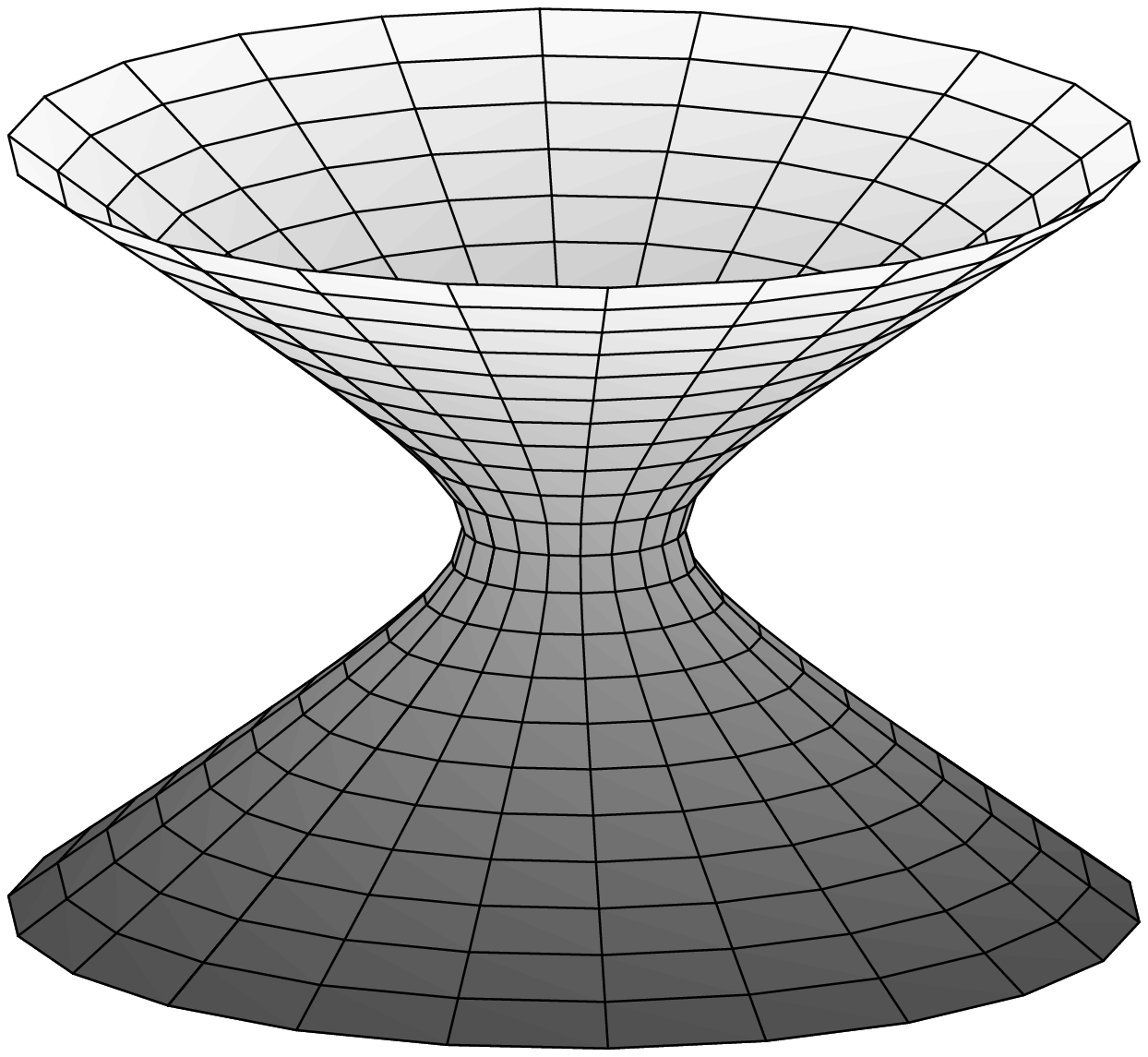}
\includegraphics[height=6cm,angle=-90]{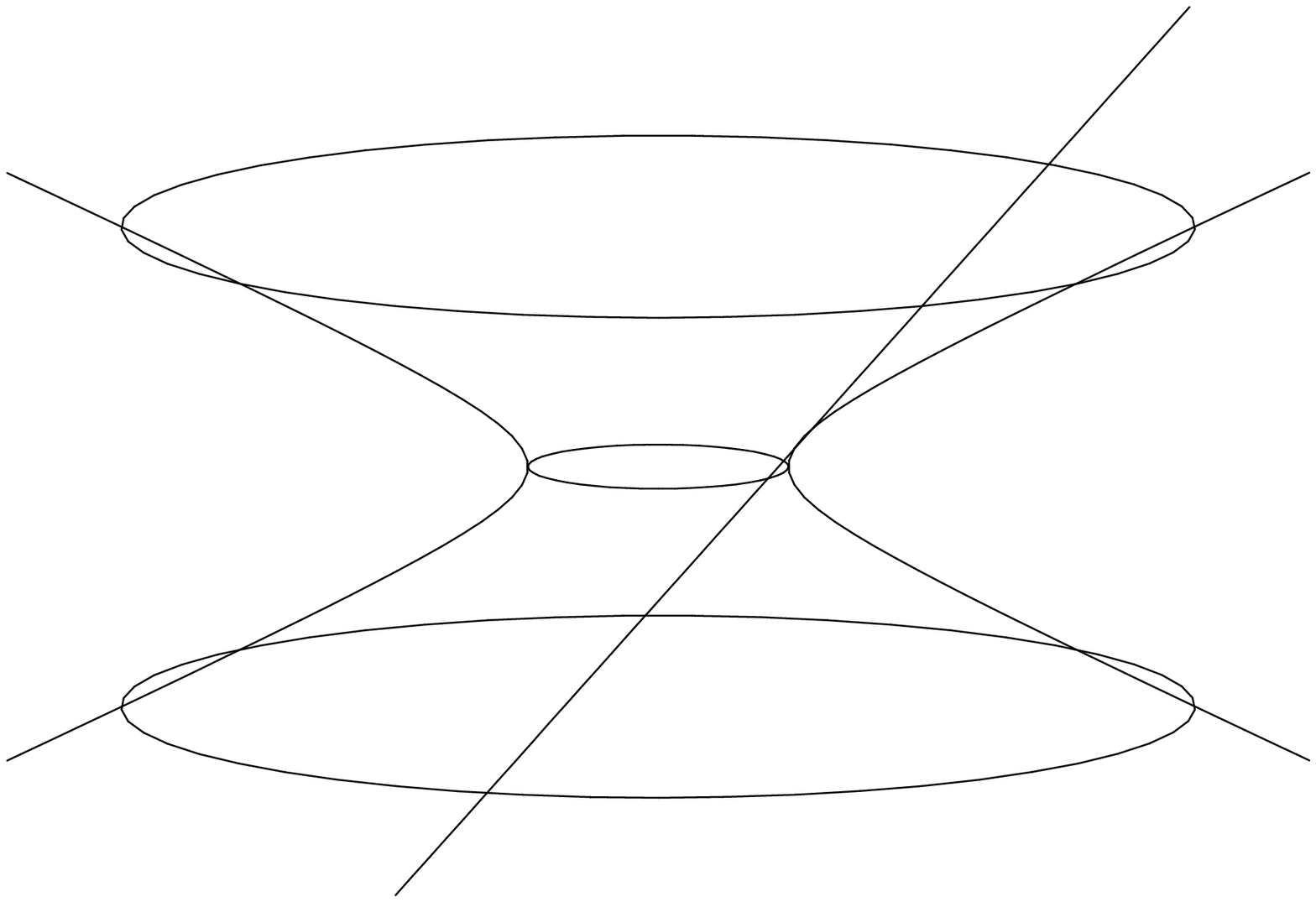}
\caption{\textit{Two-dimensional de Sitter Space}.  The
  timelike direction is vertical, and spatial sections are closed.
  The right-hand diagram shows a null geodesic, which is a straight
  line in the embedding space.}
\label{fig4}
\end{center}\end{figure}

De Sitter space is a space of constant negative curvature, forming the
Lorentzian analogue of the non-Euclidean geometry discovered by
Lobachevskii and Bolyai~\cite{brannan-cup,gap}.  Two-dimensional
non-Euclidean geometry has an elegant construction in terms of the
Poincar\'{e} disc (which was, in fact, first given by Beltrami in
1868, and later rediscovered by Poincar\'{e}~\cite{pen-road}.)  In the
Poincar\'{e} disk model, geodesics are represented as `$d$-lines' ---
circles that intersect the disc boundary at right-angles (see
figure~\ref{fig2}).  Here we develop a similar picture for
2-dimensional de Sitter space, which is sufficient to capture the key
features of the geometry.  We start with an embedding picture,
representing de Sitter space as the 2-surface defined by
\begin{equation}
T^2 - X^2 - Y^2 = -a^2,
\end{equation}
where $(T,X,Y)$ denote coordinates in a space of signature $(1,2)$ and
$a$ is a constant.  The resulting surface is illustrated in
figure~\ref{fig4}.  The figure illustrates a key property of the
entire de Sitter geometry, which is that spatial sections are closed,
whereas the timelike direction is open.  So de Sitter space describes
a closed universe that lasts for infinite time.  One can set up local
coordinate patches for which spatial sections are flat or open, but
these coordinates are not global.  These are discussed in the
following section.  Null geodesics are straight lines formed by the
intersection of the surface and a vertical plane a distance $a$ from
the timelike axis.  Despite the fact that the space is spatially
closed, the furthest a photon can travel is half of the way round the
universe.

\begin{figure}\begin{center}
\includegraphics[height=4cm]{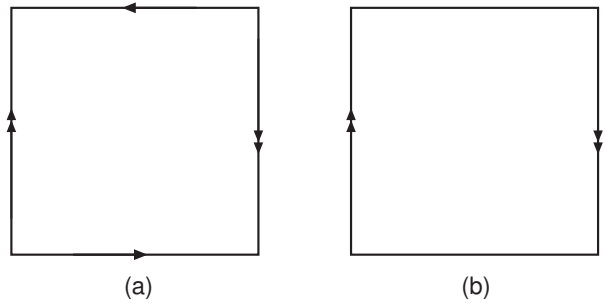}
\caption{(a) A representation of the projective plane.  Common arrows
  denote surfaces to be identified, together with the orientation.
  (b) The M\"{o}bius strip, shown for comparison.}
\label{fig3.5}
\end{center}\end{figure}

The preceding picture is the `standard' one of de Sitter space, though
it is not the one originally put forward by de
Sitter~\cite{klemm04,mcinn03}.  Further spaces of constant curvature
can be created by topological identifications, and among these is the
antipodal map on spatial sections.  In terms of embedding coordinates
$(T,X^i)$, this is the map $(T,X^i)\mapsto(T,-X^i)$.  The result has
spatial sections of the form $\Rl P^n$ as opposed to $S^n$, where $\Rl
P^n$ denotes the real $n$-dimensional projective plane.  The
projective plane can be quite hard to visualise.  For example, $\Rl
P^2$ involves taking a Mobius strip, and applying a further twisted
identification on the `long' side (figure~\ref{fig3.5}).  The
resulting manifold in non-orientable, and cannot be embedded in three
dimensions without self-intersection (it is similar to a Klein
bottle).  Of more physical relevance is $\Rl P^3$, which is
orientable, though the space contains non-contractable loops and is
equally hard to visualise.  $\Rl P^3$ is also a group manifold, so as
a background space it does have some attractive properties.  The
3-sphere $S^3$ is the group manifold of the three dimensional spin
group (and also SU(2)), and $\Rl P^3$ is correspondingly the group
manifold of SO(3).  From our perspective, the standard framework of
$S^n$ spatial sections is the one of greatest interest, though this is
largely an aesthetic judgement.  There does not appear to be any
obvious physical criteria for favouring one over the other.

\begin{figure}\begin{center}
\includegraphics[height=5cm]{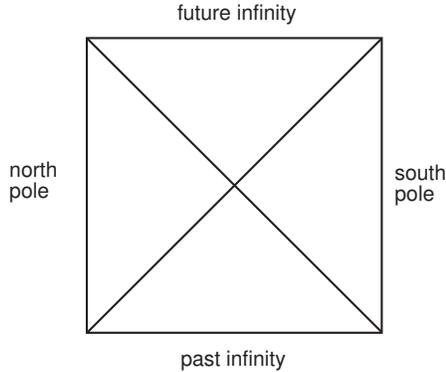}
\caption{\textit{Conformal diagram of de Sitter space.}  The vertical
  axis represents conformal time, running from $0$ to $\pi$. This
  version of the diagram is appropriate for a 4-dimensional spacetime,
  with $\theta$ and $\phi$ coordinates suppressed.}
\label{fig4.5}
\end{center}\end{figure}

\begin{figure}\begin{center}
\includegraphics[height=6cm]{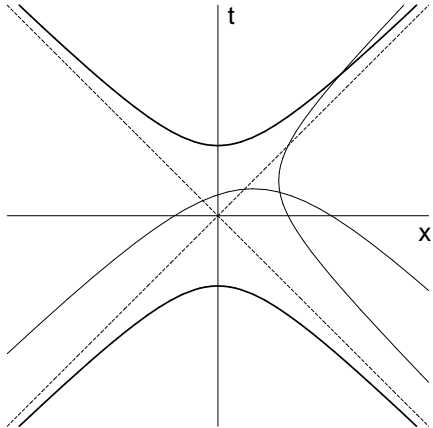}
\caption{\textit{The Lorentzian view of two-dimensional de
Sitter space}.  The boundary is defined by two hyperbolae (shown with
thick lines).  All geodesics through the origin are straight lines,
and null geodesics are always straight lines at $45^\circ$.  Two
further geodesics, one spacelike and one timelike, are also shown.
These are hyperbolae which do not pass through the origin.  The
timelike geodesic intersects the boundary in such a way that the two
tangent vectors have vanishing Lorentzian inner product.}
\label{fig6}
\end{center}\end{figure}

There are a range of conformal representations one can adopt for de
Sitter space~\cite{bir-quant}.  Probably the simplest is the
Carter--Penrose diagram, shown in figure~\ref{fig4.5}.  In this
picture the temporal coordinate represents conformal time, and the
spatial section represents the effective radial coordinate.  A
2-sphere is suppressed at each point.  If one wants to represent the
simpler 2-dimensional de Sitter space, the diagram needs to be wrapped
onto a cylinder.  While this picture is instructive, we seek a picture
that embodies some of the features of the Poincar\'{e} disk model.
So, for example, geodesics should have a simple representation in
terms of curves of constant (Minkowski) distance from some point.  To
achieve such a picture, we start by considering the spatial section at
$T=0$.  This section is a ring of radius $a$, which is mapped onto a
straight line in a $(1,1)$ Lorentzian space via a stereographic
projection.  Null geodesics from this section are represented as
$45^\circ$ straight lines in Lorentzian space.  Since geodesics from
opposite points on the ring meet at infinity, we arrive at a boundary
in the timelike direction defined by a hyperbola.  This construction
provides us with a Lorentzian view of de Sitter geometry.  Timelike
geodesics in de Sitter space are represented by hyperbolae that
intersect the boundary at a (Lorentzian) right-angle (see
figure~\ref{fig6}).

There are many fascinating geometric structures associated with de
Sitter geometry, many of which mirror those of non-Euclidean
geometry. For example, one can always find a reflection that takes any
point to the origin~\cite{anl-confcos}.  One can then prove a number
of results at the origin, where the geodesics are all straight lines,
and the results are guaranteed to hold at all points.  A similar
approach can be applied to anti-de Sitter space, except now the
diagram is rotated through $90^\circ$, as it is the timelike direction
that is formed from a stereographic unwrapping of a circle.  A
valuable mathematical tool for studying these geometries is provided
by \textit{conformal geometric algebra}~\cite{gap,Dll-sheff}, which is
essentially a Clifford algebra for the conformal space of two
dimensions higher.  So a 4-dimensional de Sitter spacetime is studied
in the Clifford algebra of a 6-dimensional space with signature (2,4).
Points, lines (including geodesics), planes and spheres are all
represented by graded multivectors in the algebra, and union and
intersection operations can be defined analogously to projective
geometry~\cite{anl-confcos}.

The metric associated with the Lorentzian view of 2-dimensional
de Sitter geometry has the conformally-flat structure
\begin{equation}
ds^2 = \frac{a^4}{(a^2+\cdx^2-\cdt^2)^2}(d\cdt^2 - d\cdx^2).
\end{equation}
This makes it clear that null geodesics must remain as straight lines
in the $\cdx$--$\cdt$ plane.  A similar line element is appropriate
for four dimensional spacetimes, but these are not commonplace in the
literature (except for the simple case of flat cosmologies).  We
finish this section by outlining the steps required to transform a
standard FRW line element in
the form
\begin{equation}
ds^2 = dt^2 - \frac{R(t)^2}{(1+k r^2/4)^2} \bigl(dr^2 + r^2 (d\theta^2
+ \sin^2(\theta) \, d\phi^2) \bigr)
\label{FRW}
\end{equation}
into the spacetime-conformal line element
\begin{equation}
ds^2 = \frac{1}{f^2} \bigl( d \cdt^2 - d\cdr^2 - \cdr^2 (d\theta^2
+ \sin^2(\theta) \, d\phi^2) \bigr) ,
\label{cFRW}
\end{equation}
where $f$ is a (dimensionless) function of $\cdt$ and $\cdr$.

The first problem to address is that the $r$ coordinate in
equation~\eqref{FRW} is assumed to be dimensionless.  To rectify this
we introduce a constant distance $\lambda$ and replace the line
element of~\eqref{FRW} with
\begin{equation}
ds^2 = dt^2 - \frac{4 \lam^2 R(t)^2}{(\lam^2 + k r^2)^2} \bigl(dr^2 +
r^2 (d\theta^2 + \sin^2(\theta) \, d\phi^2) \bigr).
\end{equation}
Here $t$, $r$, $\lam$ and $R$ all have units of distance (assuming
$c=1$). As usual, the constant $k$ is either $\pm 1$ or zero.
Comparing the angular terms in equations~\eqref{FRW} and~\eqref{cFRW}
we see immediately that
\begin{equation}
\frac{\cdr}{f} = \frac{2 \lam r R(t)}{\lam^2 +k r^2}.
\end{equation}
The coordinate transformation from $(t,r)$ to $(\cdt,\cdr)$ must
satisfy
\begin{align}
d\cdt &= f \cosh(u) \, dt + \frac{\cdr}{r} \sinh(u) \, dr \nn \\
d\cdr &= f \sinh(u) \, dt + \frac{\cdr}{r} \cosh(u) \, dr .
\end{align}
For flat cosmologies ($k=0$) we simply set $\cdt/\lam$ equal to the
conformal time $\eta$, where
\begin{equation}
\eta = \int_0^t \frac{dt'}{R(t')},
\label{defeta}
\end{equation}
and we then have $\cdr=r$ and the hyperbolic angle $u$ is set to zero.
For non-flat cosmologies it is perhaps surprising to find that $u$ is
non-zero.  That is, there is a mismatch between the conformal
coordinate frame and the cosmological frame.  It follows that the
conformal time $\eta$ is not the same as the time-like conformal
coordinate $\cdt$.

Using the integrability conditions for the coordinate
transformations, and setting the initial singularity to  $\cdt=0$, 
the following solutions are found in the three
cosmological scenarios~\cite{DL03-deS}:
\begin{align}
f &= \frac{\cdt}{R \sin(\eta)} =
g \left( \frac{2 \lam \cdt}{\lam^2 + \cdr^2 - \cdt^2} \right) \,
\frac{\cdt}{\lam} &
\mbox{closed} \nn \\
f &= \frac{2 \lam}{R} & \mbox{flat}\label{formsoff}  \\
f &=\frac{\cdt}{R \sinh(\eta)} = \bar{g} \left(\frac{2 \lam
  \cdt}{\lam^2 + \cdt^2 - \cdr^2} \right)  \,
\frac{\cdt}{\lam} & \mbox{open} \nn.
\end{align}
Here $g$ and $\bar{g}$ are functions (in general elliptic)
satisfying differential equations depending on the matter
content.

\section{A boundary condition for conformal time}

\begin{figure}\begin{center}
\includegraphics[width=5.5cm]{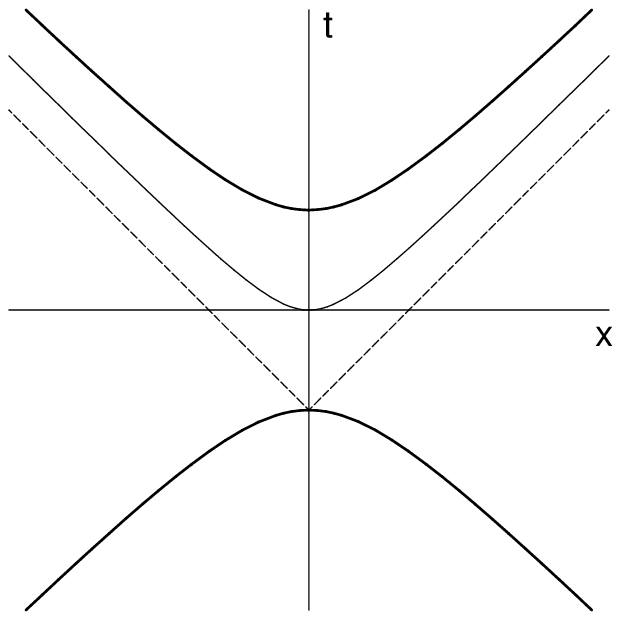}
\includegraphics[width=5.5cm]{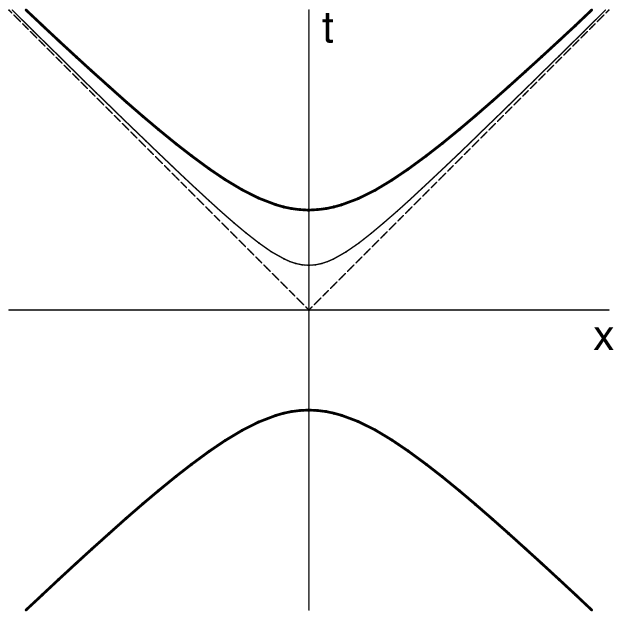}
\caption{\textit{Flat and open cosmological models}.  de
Sitter space contains sections representing both flat and open
spacetimes.  A flat spacetime (left) consists of the space contained
inside a light-cone located on past infinity.  An open spacetime fits
inside the lightcone from the origin.  Both pictures can be used to
illustrate cosmological models, with the initial singularity
represented by a hyperbolic spacelike surface.  An example of these
surfaces is shown on each diagram.}
\label{fig7}
\end{center}\end{figure}

All cosmological models are conformally flat, and can all be
interchanged via conformal transformations.  Furthermore, all models
containing a cosmological constant, and which do not recollapse, will
tend towards a de Sitter end state.  Such models should fit neatly
within the conformal diagram of figure~\ref{fig6} with future infinity
represented by the upper hyperbolic boundary.   A flat section of de
Sitter space corresponds to a region contained within a light cone
from a point located on the boundary at past infinity (see
figure~\ref{fig7}).  Surfaces of constant cosmic time are represented
as hyperbolae, any one of which can then be chosen to represent the
initial singularity.  Similarly, an open section of de Sitter space is
represented by the area inside a light-cone from a point in the middle
of the de Sitter picture.  An open $\Lam$-cosmology will
have the initial singularity located on a spacelike hyperbola.

The diagrams for flat and open cosmologies make it clear that one is
not employing the full de Sitter geometry in a symmetric
manner. Closed models, on the other hand, can be given a more natural
embedding, since their initial singularity can be represented by a
spacelike surface placed anywhere in the conformal diagram of
figure~\ref{fig6}. For all but one choice of position of the initial
singularity, the future asymptotic de Sitter state of the model will
not match onto the future infinity boundary of the conformal
diagram. However, suppose that we insist that $\cdt=0$ represents the
initial singularity. In this case the entire history of the universe
can be conformally mapped into the top half of the de Sitter
diagram. We suggest that this gives a natural boundary condition, as
the centre of the diagram is the most natural, symmetric place to
locate the big-bang singularity. (This choice also hints at a possible
pre-big bang phase, but this is not explored here.) It might be
thought that our proposal just amounts to a choice of
coordinates. However, we now demonstrate that it has physically
measurable consequences, and also that it can be recovered via an
eigenvalue-type condition on the underlying equations, hinting at its
origin as a possible quantisation condition.

One immediate physical consequence of our proposal is that any
photon emitted from the initial singularity may travel a maximum
of precisely one-quarter of the way round the universe in the
entire future evolution of the universe.  An alternative way of
saying this is that the past horizon projected back to $\cdt=0$
should cover half of the de Sitter geometry.  If we let $\phi$
denote an equatorial angle on a 3-sphere, a photon travelling
round the equator will satisfy
\begin{equation}
\frac{d\phi}{dt} = \frac{1}{R},
\end{equation}
where $t$ is cosmic time.  Traversing the entire universe corresponds
to running from $0\leq \phi \leq 2\pi$.  If a photon is only to travel
one-quarter of the way round we therefore require that
\begin{equation}
\int_0^\infty \frac{dt}{R} = \int_0^\infty \frac{dR}{R(-1 + \Lam R^2/3 +
  8\pi \rho R^2/3)^{1/2}} = \pi/2.
\label{etacond}
\end{equation}
That is, the total available conformal time is $\pi/2$.  This
constraint can also be arrived at through an alternative route
that works entirely within the conformal representation of
cosmological models~\cite{anl-confcos}. A de Sitter space centred
on $t=0=\cdt$ has the conformal line element
\begin{equation}
ds^2 = \frac{12 \lam^2}{\Lam(\lam^2 + \cdr^2 - \cdt^2)^2}  \bigl(
d\cdt^2 - d\cdx^2 - d\cdy^2 - d \cdz^2 \bigr).
\end{equation}
Clearly the only forms for $f$ in equation~\eqref{formsoff} that have
any chance of matching onto this final state are those for a closed
universe.  Furthermore, the function $g$ must satisfy
\begin{equation}
\lim_{\chi \mapsto \infty} g(\chi) = \left( \frac{\lam^2 \Lam
}{3}\right)^{1/2} \frac{1}{\chi}
\end{equation}
where
\begin{equation}
\chi = \frac{2 \lam \cdt}{\lam^2 + \cdr^2 - \cdt^2} = \tan(\eta).
\end{equation}
But since
\begin{equation}
g = \frac{\lam}{R \sin(\eta)}
\end{equation}
we must then have $R \cos(\eta)$ tending to a constant at large
times. This is only possible if $\eta$ tends to $\pi/2$, recovering
our earlier boundary condition.  This derivation is instructive in
that it reveals how the constraint can be imposed as a straightforward
boundary condition on a differential equation.  In this case, the
equation for $g(\chi)$ is
\begin{equation}
\chi^2(1+\chi^2) \left( \frac{dg}{d\chi}\right)^2 + \frac{d}{d\chi}
(g^2 \chi) = \frac{8 \pi G \lam^2 \rho}{3} + \frac{\lam^2 \Lam}{3}.
\end{equation}
The task then is to solve this equation subject to the boundary
condition that $g$ falls off as $1/\chi$ for large $\chi$.  Viewed
this way the constraint can be thought of as a `quantisation
condition' applied as the universe is formed, which one might hope
would emerge from a quantum theory of gravity.

\begin{figure}\begin{center}
\psfrag{Omm}{\footnotesize $\Omega_M$}
\psfrag{Oml}{\footnotesize $\Omega_\Lambda$}
\includegraphics[height=5cm]{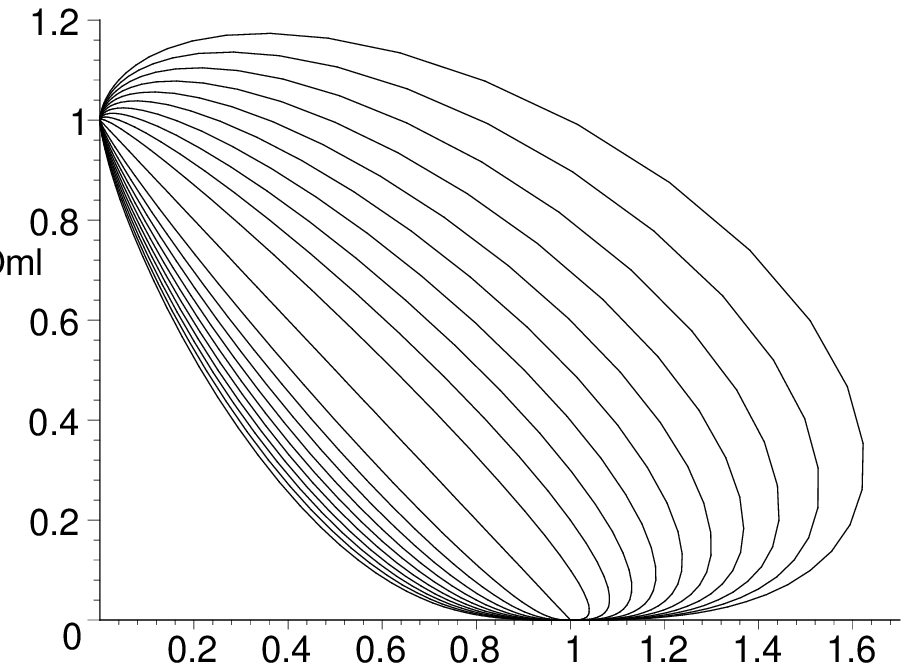}
\includegraphics[height=5cm]{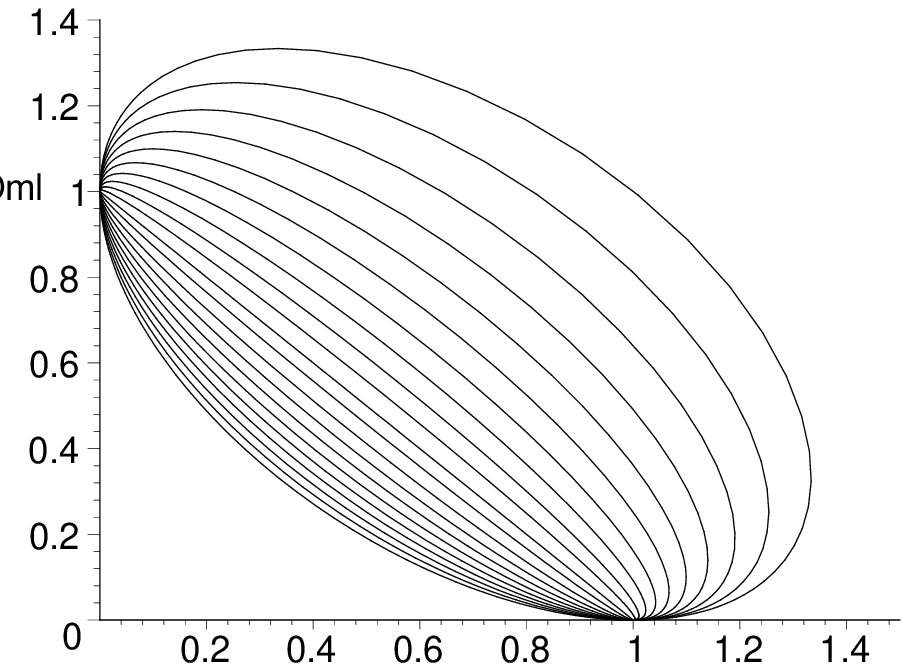}
\caption{Evolution curves in the $(\Om_M, \Om_\Lam)$
plane.  The left-hand plot is for dust, and the right-hand plot is for
radiation.  In both cases the curves converge to $\Om_\Lam=1$,
representing a late-time de Sitter phase.}
\label{fig1}
\end{center}\end{figure}

In order to understand the implications of our boundary condition, we
turn to considering flow lines in the $(\Om_M, \Om_\Lam)$ plane (see
figure~\ref{fig1}).  The plots show a shows a series of flow lines
starting from $\Om_M=1$, $\Om_\Lam=0$, refocusing around the
spatially flat case, $\Om_M+\Om_\Lam=1$.  One can show that, for a
large range of initial conditions, by the time we reach the current
value of $\Om_M \approx 0.3$ most models are not far off spatial
flatness.  This prediction contrasts with models without a
cosmological constant, where any slight deviation from the critical
density in the early universe is scaled enormously by the time we
reach the present epoch, implying that the parameters in the early
universe are highly fine-tuned.  The presence of a cosmological
constant therefore goes some way to solving the flatness problem on its
own, without the need to invoke inflation.

\begin{figure}\begin{center}
\psfrag{Omm}{\footnotesize $\Omega_M$}
\psfrag{Oml}{\footnotesize $\Omega_\Lambda$}
\includegraphics[height=6cm]{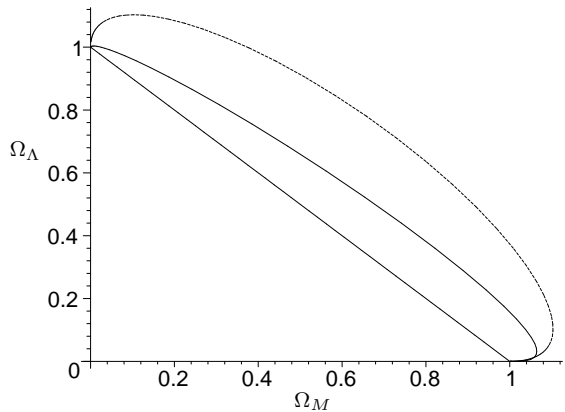}
\caption{\textit{Critical paths as predicted by the de Sitter
embedding}.  The solid line represents a matter-dominated universe,
and the broken line shows radiation, for comparison.  The straight
line is the critical case of spatial flatness.  By the time we reach
$\Om_M \approx 0.3$ the universe has been driven close to critical
density.}
\label{fig8}
\end{center}\end{figure}

Each of the flow lines in figure \ref{fig1} has a different value for
the total evolved conformal time.  Imposing a constraint on this picks
out a single preferred trajectory, resulting in the two curves shown
in figure~\ref{fig8} (one for matter-filled and one for
radiation-filled).  As the universe is expected to be matter dominated
for most of its history, the solid line in figure~\ref{fig8} is the
more physically relevant one.  Taking the present-day energy density
to be around $\Om_M = 0.3$, we see that $\Om_\Lam=$ is predicted to be
around $\Om_\Lam \approx 0.83$.  That is, a universe that is closed at
around the 10\% ratio.  Such a prediction is reasonably close to the
observed value, though it is ruled out by the most recent
experiments~\cite{WMAP}.  In order to improve the prediction, we need
to use up a greater fraction of the conformal time before we enter the
matter-dominated phase.  Such a process is also necessary to solve the
horizon problem, and the simplest means of achieving this is via an
inflationary phase.

\section{Scalar field inflation}

One can argue about the extent to which inflation really does
solve the original problems it was intended to.  For example, we
have seen that the presence of a cosmological constant alone goes
a long way to solving the flatness problem, without requiring
inflation to put the universe in an effective de Sitter phase.
But when it comes to generating structure on scales greater than
the horizon, inflation is currently the only option.  The
simplest model of inflation models the matter content as a real,
time-dependent, homogeneous massive scalar field.  For a range of
initial conditions this system shows the expected inflationary
behaviour.  The evolution equations for this model are
\begin{equation}
\dot{H} + H^2 - \frac{\Lambda}{3} + \frac{4 \pi G}{3} ( 2
\dot{\phi}^2 - m^2 \phi^2) = 0
\label{sceq1}
\end{equation}
and
\begin{equation}
\ddot{\phi} + 3 H \dot{\phi} + m^2 \phi = 0.
\label{sceq2}
\end{equation}
Given a solution to the pair of equations~\eqref{sceq1}
and~\eqref{sceq2}, a new solution set is generated by scaling with a
constant $\sigma$ and defining
\begin{equation}
H'(t) = \sigma \Hbar(\sigma t), \quad \phbar' (t) =
 \phbar(\sigma t), \quad m' =  \sigma m, \quad \Lbar' =
\sigma^2 \Lbar.
\label{scltrf}
\end{equation}
This scaling property is valuable for numerical work, as a range of
situations can be covered by a single numerical integration.  A range
of physical quantities are invariant under changes in scale, including
the conformal time $\eta$ as can be seen from equation~\eqref{defeta}.
The scaling property does not survive quantisation, however, so has to
be employed carefully when considering vacuum fluctuations.

The initial conditions for equations~\eqref{sceq1}
and~\eqref{sceq2} are usually set at the start of the
inflationary period, where they are viewed as arising from some
form of quantum gravity interaction. But in order to apply our
boundary condition we need to track the equations right back to
the initial singularity, as this is the only way that we can keep
track of the total conformal time that has elapsed.  Evolving the
inflationary equation backwards in time is entirely justified, as
we do not expect quantum gravity to play a role until much
earlier in the history of the universe.  Inflation on its own
does not eliminate singularities from
cosmology~\cite{brand01,borde03}.  Furthermore, by specifying
initial conditions around the singularity, the states of the
fields at the onset of inflation are fully determined by a pair
of parameters, making the model highly predictive.

As the universe emerges from the big bang the dominant behaviour of
$H$ is to go as $1/(3t)$.  Equation~\eqref{sceq1} then implies that
$\phi$ must contain a term going as $\ln(t)$.  But this in turn
implies that $H$ must also contain a term in $t\ln(t)$, in order to
satisfy equation~\eqref{sceq2}.  Working in this manner we conclude that
a series expansion in powers of $t$ and $\ln(t)$ is required to describe the
behaviour around the singularity.  At this point it is convenient to
define the dimensionless variables
\begin{equation}
u = \frac{t}{t_p}, \qquad \mu = \frac{m}{m_p}
\end{equation}
with $t_p$ and $m_p$ the Planck time and mass respectively.  The
series expansion about the singularity at $t=0$ can now be written
\begin{equation}
\Hbar(u) = \frac{1}{t_p}\sum_{i=0}^\infty H_i(u) \ln^i(u), \qquad
\phbar(u) = \frac{1}{l_p} \sum_{i=0}^\infty \phi_i(u) \ln^i(u) ,
\end{equation}
which ensures that the expansion coefficients are all dimensionless.
Substituting these into the two evolution equations, and setting each
coefficient of $\ln(u)$ to zero, we establish that
\begin{align}
H_1 &= -u \diff{H_0}{u} - u H_0^2 + \frac{u \Lbar}{3} - \frac{8
\pi u }{3 } \left(\diff{\phi_0}{u} \right)^2 - \frac{16 \pi \phi_1}{3 }
\diff{\phi_0}{u} \nn \\
& \quad - \frac{8 \pi \phi_1^2}{3 u} + \frac{4 \pi \mu^2 u
\phi_0^2}{3} ,
\end{align}
with further algebraic equations holding for $H_2, \phi_2$, $H_3,
\phi_3$, and so on.  So, by specifying $H_0$, $\phi_0$ and $\phi_1$,
all the remaining terms in the solution are fixed.  The aim now is to
choose the three input functions to ensure that successive terms in
the series get progressively smaller.  This provides just the right
number of equations to specify all coefficients, save for two
arbitrary coefficients in $\phi_0$.  This results in a series
expansion controlled entirely by two arbitrary constants, which is the
expected number of degrees of freedom once we have fixed the
singularity to $t=0$.  In order to generate curvature it turns out
that the input functions need to be power series in $u^{1/3}$, which
ensures that the scale factor goes as $u^{1/3}$ at small times.  The
series solution is only required to find suitable initial conditions
for numerical evolution, so only the first few terms are required.
Expanding up to order $u^{5/3}$ we find that
\begin{align}
H_0 &= \frac{1}{3u} + \frac {32\sqrt {3\pi }}{27}  b_4 u^{1/3}
\nn \\
& \quad + \left( \frac {2 \mu^2 }{81} + \frac{\Lbar}{3} +
\frac{4 \pi}{3} \mu^2 b_0^2+  \frac {4 \sqrt{3 \pi}}{27} {\mu}^{2}
b_0 \right) u - \frac{6656 \pi b_4^2 }{891} u^{5/3}, \\
\phi_0 &= b_0  + b_4 u^{4/3} - \frac {118 \sqrt {3 \pi} b_4^2}{99}
u^{8/3} \nn \\
& \quad
- \frac {1}{1296 \pi} \left( 11 \sqrt {3 \pi} \mu^{2}
- 54 \sqrt {3 \pi} \Lbar - 216 \sqrt {3}\, \pi^{3/2} \mu^2 b_0^2 + 36
\pi \mu^2 b_0 \right) u^2
\end{align}
and
\begin{equation}
\phi_1 = - \sqrt{\frac{1}{12 \pi}} -{\frac {\mu^2}{216 \pi}} \left
(-\sqrt {3 \pi } + 36  \pi b_0\right ) u^2 .
\end{equation}
Under scaling the three key parameters in the model transform as $\mu
\mapsto \sigma \mu$, $b_0 \mapsto b_0$, and $b_4 \mapsto \sigma^{4/3}
b_4$.  This scaling transformation for $b_4$ is entirely as expected,
given that it is the coefficient of $u^{4/3}$ in the series for
$\phi_0$.  It follows that the quantity $b_4/\mu^{4/3}$ is scale
invariant.

\begin{figure}\begin{center}
\includegraphics[width=6cm,angle=-90]{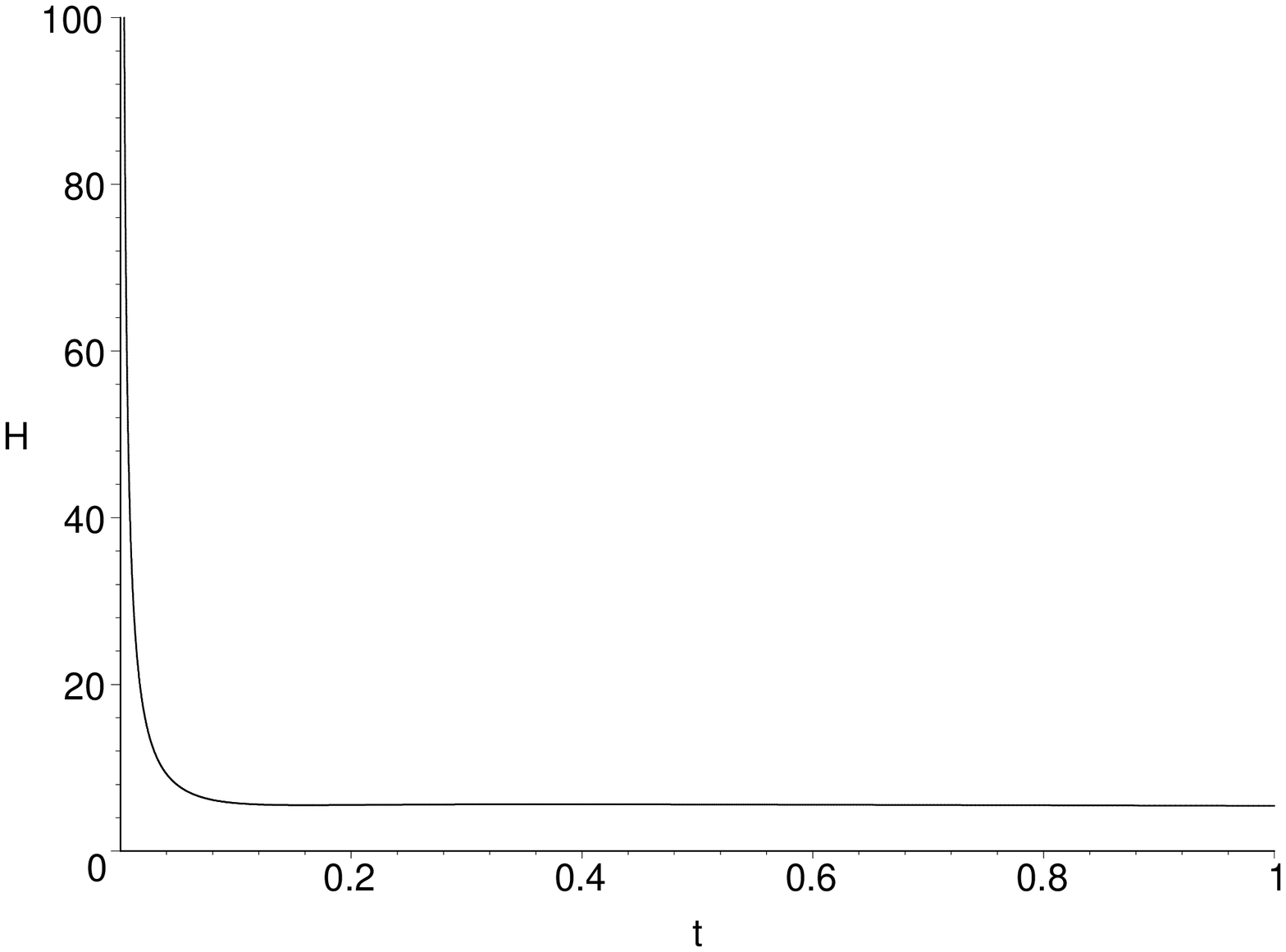}
\includegraphics[width=6cm,angle=-90]{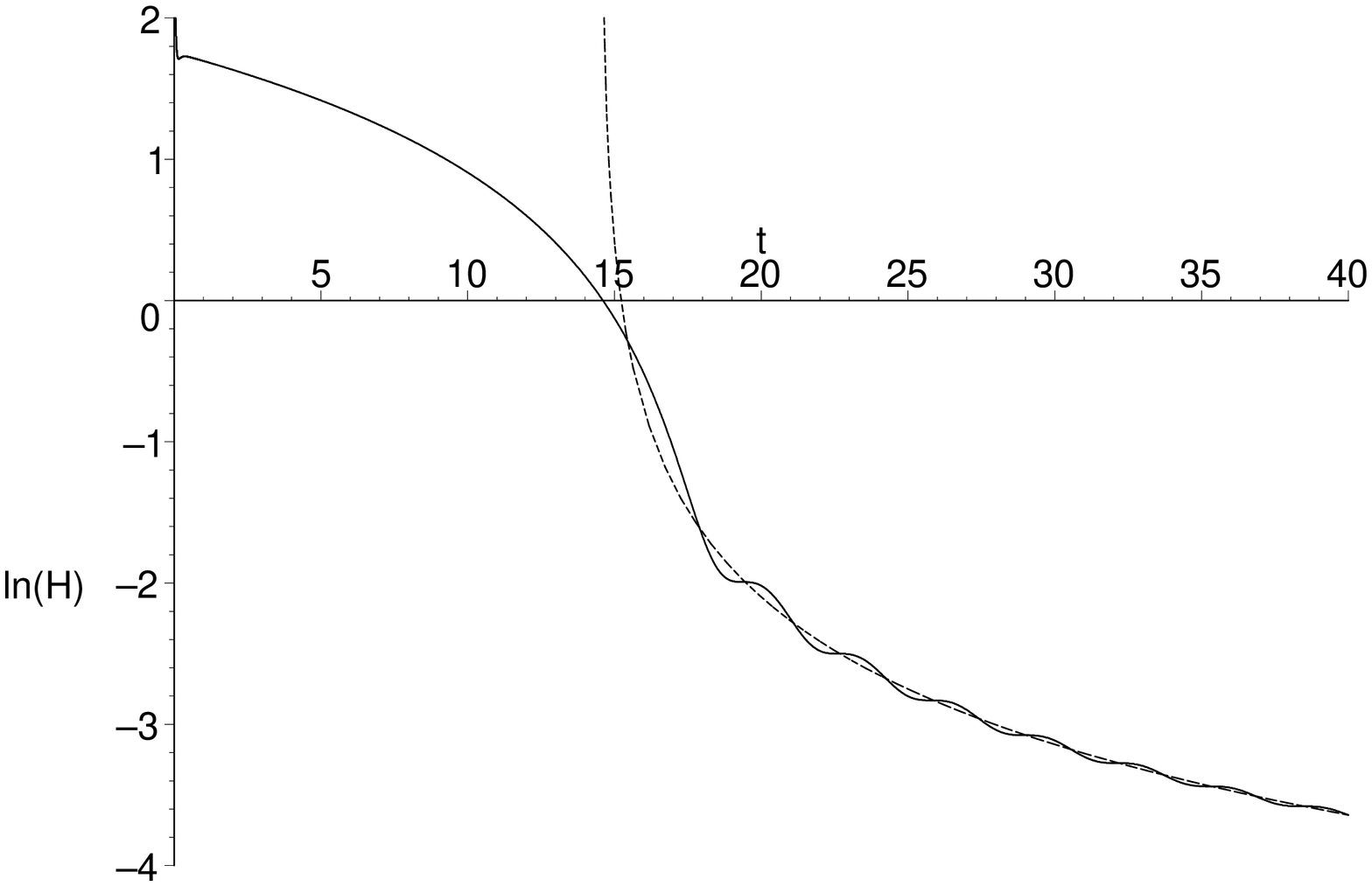}
\caption{\textit{Hubble function entry into and exit from the
inflationary regime.}  The Hubble function emerges from the big bang
going as $1/3t$.  As $H$ falls it quickly enters the inflationary
region, during which it falls linearly.  As the inflationary era ends
the density and pressure start oscillating around values for a
matter-dominated cosmology.  The effective singularity for this dust
cosmology is displaced from $t=0$.  The broken line plots the natural
log of $2/3(t-14.56)$.  Time $t$ is measured in units of $t_p$, and
$H$ in units of $t_p^{-1}$.  The input parameters were set to
$b_0=2.48$, $b_4=-0.51$ and $m=m_p$ ($\mu=1$).}
\label{fHfrom0}
\end{center}\end{figure}

The fact that $b_4$ controls the curvature can be seen from
equation~\eqref{curv} which, to leading order, yields
\begin{equation}
\frac{R}{l_p} = \frac{1}{\mu }\left( \frac{2187}{12544 \pi}
\right)^{1/4} \left( - \frac{\mu^{4/3}}{b_4} \right)^{1/2} (\mu
u)^{1/3} + \cdots.
\label{Rnear0}
\end{equation}
Clearly, the arbitrary constant $b_4$ must be negative for a closed
universe.  The terms on the right-hand side of equation~\eqref{Rnear0}
are all scale invariant, apart from the first factor of $\mu^{-1}$.
In figure~\ref{fHfrom0} we illustrate Hubble function entry into the
inflationary regime for typical values of $b_0$ and $b_4/\mu^{4/3}$ of
interest.  We see that the onset of inflation corresponds to $\mu u
\approx 0.1$.  In order to generate perturbations consistent with
observation the scalar field must have a mass of the order of
$10^{-6}m_p$.  It follows that the onset of inflation occurs at a time
of around $10^5$ Planck times.  The radius of the universe at the
onset of inflation is then given approximately by
\begin{equation}
R \approx \frac{0.2}{\mu} l_p,
\end{equation}
and so is of order $10^5$ Planck lengths.  Inflation therefore starts at
an epoch well into the classical regime.  Quantum gravity effects
would be expected to be relevant when the radius of the universe is of
the order of the Planck scale, which occurs when $u \approx \mu^2$ and
is well before any inflationary period (for physical values of $\mu$).
We are therefore quite justified in running the evolution equations
back past the inflationary regime, and right up to near the initial
singularity.  It is only when $R=l_p$ that the equations will break
down, and we would look to quantum gravity to explain the formation of
the initial, Planck-scale sized universe.  Indeed, we would argue more
strongly that we \textit{have} to run the equations backwards in time
to well before the start of inflation before reaching an epoch where
new physics could be expected to enter the problem.

As the universe is described by a 3-sphere of radius $R$,
the total volume of the universe is given to leading order by
\begin{equation}
V = 2 \pi^2 \left( \frac{2187}{12544 \pi} \right)^{3/4}
\frac{l_p^3}{(-b_4)^{3/2}} u + \cdots.
\end{equation}
Following from this, an interesting calculation we can perform in a
closed universe is to find the total energy contained within it in the
scalar field.  By integrating the energy density we find that
\begin{equation}
E_{\mbox{tot}} = \frac{\pi}{12} \left(\frac{2187}{12544 \pi}
\right)^{3/4} \frac{1}{(-b_4)^{3/2}} \frac{\hbar}{t} + \cdots \approx 0.03
\left( \frac{-\mu^{4/3}}{b_4} \right)^{3/2}  \frac{\hbar}{\mu^2 t},
\label{Etot0}
\end{equation}
and when the scale-invariant quantity $b_4/\mu^{4/3}$ is or order
unity (as required for physical models) we have
\begin{equation}
E_{\mbox{tot}}  \approx \frac{0.03}{\mu^2}  \frac{\hbar}{t}.
\end{equation}
This tell us that the action $Et$ is very large, when measured in
units of $\hbar$.  This is reassuring, as it means we are justified in
treating the universe on the whole as a classical object.

The plots in figure~\ref{fHfrom0} illustrates the general behaviour of
the Hubble function $H$.  As the universe emerges from the big bang
the energy density in the scalar field is dominated by the
$\dot{\phi}$ term, and the field behaves as if it is massless.  It
follows that $H$ initially falls as $1/(3t)$.  But once $H$ has fallen
sufficiently far we enter a region in which $m^2 \phi^2$ starts to
dominate over $\dot{\phi}^2$.  These are suitable initial conditions
for the universe to enter an inflationary phase.  By varying $b_0$,
$b_4$ and $\mu$ we control both the values of the fields as we enter
the inflationary period, and how long the inflationary period lasts.
The cosmological constant plays no significant role in this part of
the evolution. The dynamics displayed in this figure is quite robust
over a range of input parameters.  A significant point here is that we
enter the inflationary regime from a region of high $H$, as opposed to
the $H\approx 0$ value favoured in some models of quantum cosmology.
We therefore never enter the regions of parameter space where the
chaotic evolution noticed by Page~\cite{page84} and Cornish \&
Shellard~\cite{cor98} is relevant.

\begin{figure}\begin{center}
\includegraphics[width=5cm,angle=-90]{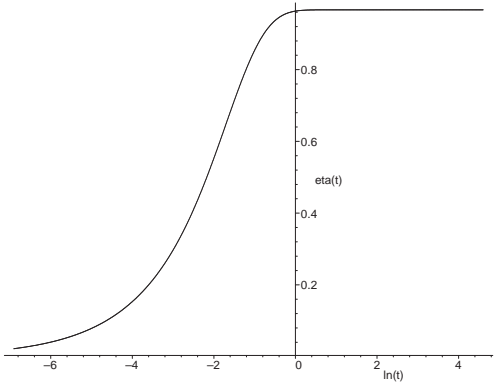}
\caption{\textit{Evolution of the conformal time $\eta(t)$ as
a function of $\ln(t)$.} The parameters for the model are as given in
figure~\ref{fHfrom0} and again $t$ is measured in units of the Planck
time.  This plot sets $\mu=1$, so the time variable must be scaled to
correspond to more physical values of $\mu \approx 10^{-6}$.}
\label{fetaevol}
\end{center}\end{figure}

The typical behaviour as we exit the inflationary region is also
shown in figure~\ref{fHfrom0}.  The end of inflation is
characterised by $\Hbar \approx \mu$.  Beyond this point the
scalar field enters an oscillatory phase, with the time-averaged
fields satisfying the conditions for a simple matter-dominated
cosmology.  In this case $H$ can be approximated by a curve going
as $2/3(t-t_0)$, describing a dust model with a displaced origin.
The universe then appears as if it has been generated by an
`effective' big bang at a later time. Around this time we expect
reheating effects to start to dominate, so that in reality the
universe must pass over to a radiation-dominated era.  But the
naive `effective big bang' concept is useful for extracting some
qualitative predictions from our model~\cite{DL03-deS}.
Figure~\ref{fetaevol} shows the evolution of conformal time for
the model depicted in figures~\ref{fHfrom0}.  As the universe
emerges from the initial singularity, $\eta$ grows at $t^{2/3}$.
But once the inflationary region is entered, $R(t)$ starts to
increase rapidly.  So the conformal time, which involves the time
integral of $R^{-1}$, quickly saturates.  So if $\eta$ has not
reached $\pi/2$ before $R$ has inflated significantly, the
universe will have to exist for an extremely long time to reach
the boundary value of $\pi/2$. Similarly, in this simple picture,
we can obtain the prediction for $\Lambda$ given in equation
(\ref{eqn:for-Lambda}) (for details see~\cite{DL03-deS}). We note
with the number of e-folds $N \approx 46$, then $\Lambda \sim
10^{-122} l_{\rm pl}^{-2}$, which is in exactly the right ball
park!

\section{The power spectrum}

\begin{figure}\begin{center}
\includegraphics[height=8cm,angle=-90]{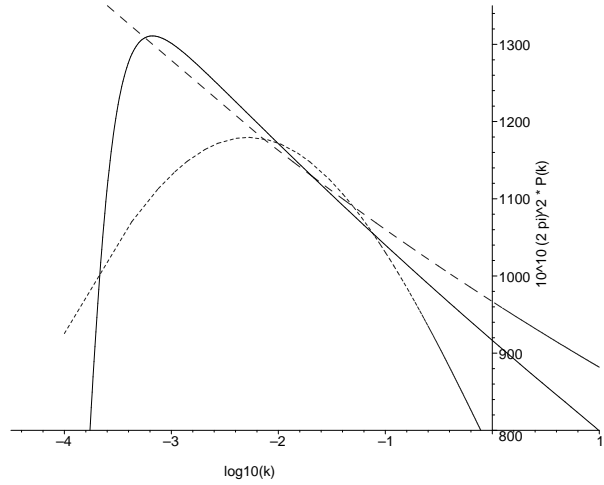}
\caption{\textit{Comparison of the scalar power spectrum of
curvature perturbations, $\clp_\clr(k)$, with
    power law models.}  The function plotted is $10^{10} (2 \pi)^2
    \clp_\clr(k)$ as a function of $\log_{10}(k)$, assuming $h=0.65$.
    The solid line represents the numerical predictions from our
    model.  The long dashes represent the best fit power law
    ($n_s=0.96$) and the short dashes are the WMAP running spectral
    index best fit.  Notice that the vertical scale runs from 800 to
    1300, so the differences are slightly exaggerated.  }
\label{sclpow2}
\end{center}\end{figure}

In~\cite{DL03-deS} we give an account of how to extract the power
spectrum of primordial fluctuations, using the various
approximations that have become standard in the literature. As an
example of the type of results achieved, we show in figure
\ref{sclpow2} a comparison of our computed scalar power spectrum
with the WMAP best fit power law and running spectral index fits.
The vertical normalisations for these fits have been chosen
arbitrarily, since it is the shape of the spectrum at large $k$
which is of greatest interest here. The running spectral index
graph is interesting in that it suggests that this model is
attempting to emulate both the cutoff at low $k$, as well as the
reduced power at large $k$, which occurs in our model. The latter
occurs as a result of computing the perturbation spectrum more
accurately than usual in this regime, and is not a specific
feature of using a closed universe model.

\begin{figure}\begin{center}
\includegraphics[height=10cm,angle=-90]{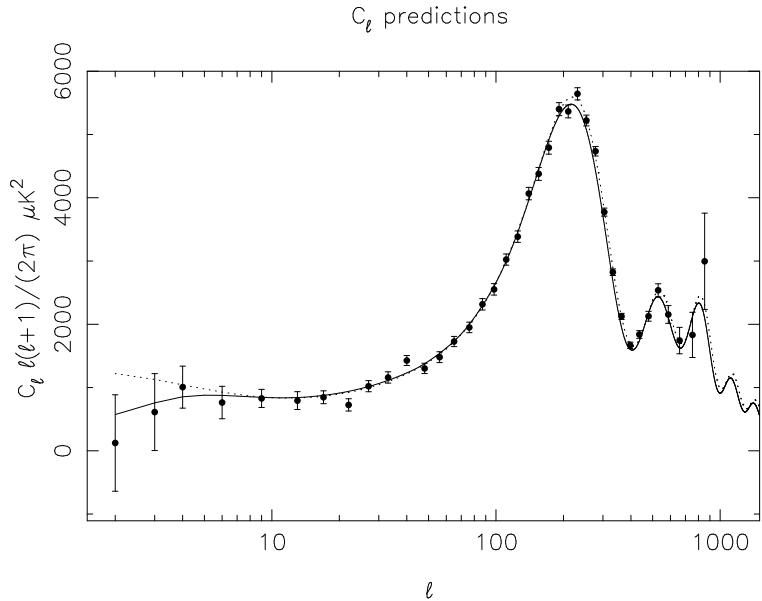}
\caption{\textit{CMB power spectrum for a model with $\Omega_0 =
1.04$} The parameters are discussed in the text. The experimental
points shown with $1\sigma$ error bars are the WMAP
results~\cite{WMAP} and the dashed curve corresponds to the best
fit $\Lambda$ CDM power law CMB power spectrum as distributed in
the WMAP data products.} \label{fp04}
\end{center}\end{figure}

In figure~\ref{fp04} we show a predicted CMB power spectrum from our
model, in comparison with the WMAP results and the prediction from a
strictly power law initial spectrum. This model has $\Om_0=1.04$, so
is consistent with WMAP at the $1\sigma$ level.  We also take
$\Omega_{\rm b}h^2=0.0224$, $h=0.60$ and $\Omega_{\rm cdm} h^2 =
0.110$.  Together these yield a value of $\Omega_{\rm cdm}=0.306$,
which is reasonable, although of course the $H_0$ value is rather
low. It can be seen that our predicted CMB spectrum is in much better
agreement with the WMAP results at low $\ell$ than the predictions
from the primordial power law.  This could be taken as good evidence
for our model, except that it is necessary to check the correctness of
the approximations we have used in calculating the perturbation power
spectrum. This work is currently in progress, but we end this paper
with a brief discussion of some of the technical problems involved.

The starting point for calculating the scalar curvature spectrum
in a simple flat model is to write the linearised perturbed
action in the form~\cite{muk92}
\begin{equation}
\del_2S = \frac{1}{2} \int d\eta \,d^3x \left( v'^2 - \eta^{ij}
v_{,i} v_{,j} + \frac{z''}{z} v^2 \right).
\end{equation}
Here $v$ is a gauge-invariant combination of matter and metric
perturbations, dashes denote the derivative with respect to conformal
time, and $z$ is defined by $z = \phi_0'/H$, where $\phi_0$ is the
unperturbed field.  By working with this action the entire problem is
reduced to one of analysing a scalar field with a time-dependent mass
term.  This approach generalises to the non-flat case, although here
the definition of $z$ becomes more complicated~\cite{zhang03}. Also,
for closed models, the mode expansions necessary to carry out
quantisation have to be performed using spatial sections given by the
3-sphere $S^3$, so that the `comoving wavenumber' $k$ takes on integer
values, with $k=3$ its lowest (non-gauge) value.  The mode equations
found this way are very complicated, but retain the general form
\begin{equation}
v_{k}'' + \left(k^2 - f(\eta,k)\right) v_{k} = 0,
\end{equation}
where in the flat case $f(\eta,k)$ would be $z''/z$, and in all cases
$f(\eta,k)$ is calculable from knowledge of the background evolution.

The challenge now is to find suitable `quantum initial
conditions' so that after evolution through inflation, the
variables $v_k$ can be used to find the perturbation spectrum. To
achieve this involves a mode decomposition of $v_{k}$ into
positive and negative frequencies.  The standard way to
approach this is to assume that in the asymptotic past the
background is either Minkowski or de Sitter, so that one knows
the correct vacuum to chose.  But this is clearly inappropriate
here, as looking back in time we encounter the initial
singularity.

The question of how to proceed in the absence of any asymptotic notion
of the vacuum state has been widely discussed. One simple approach is
provided by Hamiltonian diagonalisation, where on each time slice
modes are decomposed into positive and negative energy states by the
Hamiltonian operator.  But this technique tends to overestimate the
particle production rate. A clear way to proceed was developed by
Parker and Fulling~\cite{par69,ful-qft} and introduces the concept of
an adiabatic vacuum. The application of this to the present case is
complicated, but feasible, and initial results support a low-$k$
cutoff, though with less pronounced effects than we have found using
the standard approximate techniques. This will be the subject of a
future publication.


\end{document}